\documentclass[twocolumn, prd, preprintnumbers, amsmath,
  amssymb, showpacs, floatfix]{revtex4-1}

\usepackage{mathbbol}

\usepackage{color}
\usepackage{setspace}
\usepackage{latexsym}
\usepackage{verbatim}

\usepackage{graphicx}

\usepackage{subfigure}
\usepackage{rotating}
\usepackage{bm}
\usepackage{tikz}

\definecolor{myred}{rgb}{1.00,0,0}

\definecolor{myblue}{rgb}{0,0,1.00}

\usepackage{ulem}
\renewcommand{\emph}[1]{{\tt #1}}

\newlength{\twopicwidth} 
\newlength{\threepicwidth} 

\newcounter{ecutleft} 
\newcounter{ecutright} 
\newcounter{ecutbottom} 
\newcounter{ecuttop} 

\newcounter{dcutleft} 
\newcounter{dcutright} 
\newcounter{dcutbottom} 
\newcounter{dcuttop} 

\newcounter{scutleft} 
\newcounter{scutright} 
\newcounter{scutbottom} 
\newcounter{scuttop} 

\newcommand{\msub}[1]{\ensuremath _{\mbox{\scriptsize #1}}}

\title{Article draft}
\begin{document}
\title{Poisson - Random Matrix transition in the QCD Dirac spectrum}

\author{Tam\'as G.\ Kov\'acs}\affiliation{Institute for Nuclear Research of
  the Hungarian Academy of Sciences \\ H-4026 Debrecen, Bem t\'er 18/c, Hungary}
\author{Ferenc Pittler}\affiliation{Department of Physics, University of
  P\'ecs, H-7624 P\'ecs, Ifj\'us\'ag \'utja 6, Hungary}

\begin{abstract}
At zero temperature the lowest part of the spectrum of the QCD Dirac operator
is known to consist of delocalized modes that are described by random matrix
statistics.  In the present paper we show that the nature of these eigenmodes
changes drastically when the system is driven through the finite temperature
cross-over. The lowest Dirac modes that are delocalized at low temperature
become localized on the scale of the inverse temperature. At the same time the
spectral statistics changes from random matrix to Poisson statistics. We
demonstrate this with lattice QCD simulations using $2+1$ flavors of light
dynamical quarks with physical masses. Drawing an analogy with Anderson
transitions we also examine the mobility edge separating localized and
delocalized modes in the spectrum. We show that it scales in the continuum
limit and increases sharply with the temperature.
\end{abstract}
\maketitle

\section{Introduction}

The spectrum of the QCD quark Dirac operator is a quantity not directly
accessible by experiments but it contains essential physical information
concerning the behavior of strongly interacting systems. The most well-known
example of that is the Banks-Casher relation implying that a non-vanishing
Dirac spectral density at zero indicates the spontaneous breaking of chiral
symmetry \cite{Banks:1979yr}. Another prominent example is that in the
intermediate volume so called ``epsilon-regime'' the lowest part of the Dirac
spectrum is described by Random Matrix Theory (RMT) which makes it possible to
extract the low-energy constants of chiral perturbation theory from the
eigenvalues of the Dirac operator (see e.g.\ \cite{Verbaarschot:2000dy} and
references therein). More recently the Dirac spectrum has also been used to
determine the mass anomalous dimension of QCD-like theories with many fermions
\cite{DeGrand:2009et}-\cite{Cheng:2011ic}.

Since at low-temperature the low-lying Dirac spectrum has been so extensively
studied it is surprising how little is known about it in the high-temperature
regime. The only solid piece of information above the finite temperature
cross-over is that since chiral symmetry is restored there the spectral
density of the Dirac operator should vanish at zero. In principle Random
Matrix Theory also has predictions for the spectral statistics at such a
``soft edge''. However, unlike at zero temperature, at high temperature a
priori there is no reason to believe that the QCD Dirac spectrum is described
by this edge RMT statistics. Indeed, attempts to verify this numerically did
not produce fully convincing results \cite{Farchioni:1999ws,Damgaard:2000cx}.

It turns out that random matrix statistics is only one of two possible
extremes concerning the eigenvalue statistics. It corresponds to the case of
completely delocalized eigenvectors occurring only if typical (gauge field)
fluctuations can easily mix eigenmodes nearby in the spectrum. The other
extreme possibility is localized eigenmodes that cannot be mixed by typical
fluctuations. In that case the spectrum consists of independent eigenvalues
obeying Poisson statistics. Many examples of both types of behavior in large
linear systems are known both in the mathematics and in the physics
literature. In fact the Bohigas-Giannoni-Schmit conjecture asserts that
quantum systems whose classical counterparts are chaotic exhibit random matrix
type spectra whereas integrable systems after quantization have Poisson-type
spectra \cite{Bohigas:1983er}. An example where the same physical system,
depending on the circumstances, can exhibit both types behavior is Anderson
localization \cite{Anderson:1958vr}. In that case the transition of single
electron states from delocalized ones described by RMT statistics to localized
states with Poisson statistics is driven by impurities in the crystal lattice.

Already a long time ago the idea was put forward that such a transition might
also occur in QCD at finite temperature \cite{Halasz:1995vd}. Later on
numerical studies concluded that it is not the case and the Dirac spectrum is
described by RMT even above the finite temperature deconfining and chiral
transition \cite{Pullirsch:1998ke}. This conclusion, however, was based on the
study of full Dirac spectra. It is known that in the case of the Anderson
transition for weak disorder only the states along the band edge become
localized and states deep inside the band can still remain
delocalized. Therefore a statistical analysis of full spectra might not reveal
a localization transition occurring only along the band edge.

With this additional insight the idea of a localization-delocalization
transition in QCD at $T_c$ was revived sometime later
\cite{GarciaGarcia:2004hi}. Both instanton liquid calculations
\cite{GarciaGarcia:2005vj} and lattice QCD simulations were
\cite{GarciaGarcia:2006gr} done to provide evidence that such a transition
occurs at $T_c$. However, these calculations were performed around the
critical temperature and therefore it was not possible to see clear Poisson
statistics in the spectrum. For that lattice simulations were needed well
above $T_c$ where localized modes are fully developed. Indeed, further support
for localization in QCD was obtained by a detailed demonstration that the
lowest two eigenvalues of the overlap Dirac operator in quenched SU(2) gauge
theory obey Poisson statistics \cite{Kovacs:2009zj}. Finally in the same
system a clear transition was observed in the spectrum of the staggered Dirac
operator from localized Poisson modes to delocalized RMT modes
\cite{Kovacs:2010wx}. The picture emerging from these studies is that above
the finite temperature transition the lowest part of the Dirac spectrum
consists of localized modes obeying Poisson statistics. Higher up in the
spectrum there is a cross-over to delocalized modes described by random matrix
statistics. In the meantime, independently, other groups also observed the
tendency of low modes to become localized above $T_c$ \cite{Gavai:2008xe},
although the connection to an Anderson-type transition was not made by them. A
useful account of the spatial structure of low Dirac modes and their
localization properties can also be found in \cite{deForcrand:2006my}.

So far all the direct evidence for the transition came from quenched SU(2)
simulations. In the present paper we study the question whether full QCD with
physical light dynamical quarks also exhibits delocalized Dirac modes above
$T_c$. The question is non-trivial since light dynamical quarks suppress the
lowest quark modes through the quark determinant in the action. Nevertheless
we find that localization also occurs in full QCD with quarks of physical
masses. We demonstrate this by presenting the results of lattice QCD
simulations with $N_f=2+1$ flavors of dynamical staggered quarks. We also
study how the location of the transition within the spectrum depends on the
physical temperature.

The paper is organized as follows. At first in Section \ref{se:Sd} we
summarize the technical details of our QCD lattice simulations. In Section
\ref{se:AtitDs} we describe the analogy between the QCD
delocalization-localization transition and the Anderson transition. Here we
analyze in detail the unfolded level spacing distribution that can be used to
distinguish between localized and delocalized modes. We show that in larger
spatial volumes the transition becomes sharper and most likely in the
thermodynamic limit it becomes a genuine phase transition. Also in this
Section we show how to compute the mobility edge separating localized and
delocalized states in the spectrum and analyze localization in terms of the
participation ratio of eigenmodes. In Section \ref{se:Clar} we study how the
transition scales in the continuum limit. In particular, we show that the
localization length of localized states is always smaller than the inverse
temperature. We also demonstrate that the properly renormalized mobility edge
scales in the continuum limit and investigate its temperature dependence in
the temperature range $1.7T_c < T < 5 T_c$. Finally, in Section \ref{se:C} we
summarize our results and indicate further questions.

\section{Simulation details}
    \label{se:Sd}

At first we summarize the details of our lattice simulations.  We use the
Symanzik improved gauge and and the two level stout smeared staggered fermion
action of Ref.\ \cite{Aoki:2005vt} with $N\msub{f}=2+1$ flavors.  We take the
simulation parameters from the work of the Budapest-Wuppertal collaboration
who determined the lattice spacing from the kaon decay constant $f_K$ and set
the bare quark masses by requiring the pion and kaon masses to be equal to
their physical value \cite{Borsanyi:2010cj}. The bare parameters we used and
the corresponding lattice spacings are summarized in
Table~\ref{tab:bare_parameters}.

\begin{table}
 \begin{ruledtabular}
 \begin{tabular}{llll}
   $\beta$   & $m\msub{ud}$ & $m\msub{s}$ & $a$(fm) \\ \hline
   3.75    & 0.001786     & 0.05030  & 0.125    \\
   3.938   & 0.001172     & 0.03300  & 0.082     \\  
   4.08477 & 0.000836     & 0.02354  & 0.062     \\    
 \end{tabular}
 \end{ruledtabular}
 \caption{\label{tab:bare_parameters} The bare parameters we used in the
   lattice simulations; the inverse gauge coupling ($\beta$), the light quark
   mass ($m\msub{ud}$), the strange quark mass ($m\msub{s}$) and the
   corresponding lattice spacing ($a$).}
\end{table}

To explore the dependence of the localized-delocalized mode transition on the
temperature and the lattice spacing we performed the simulations at three
different lattice spacings $a=0.06, 0.082$ and $0.125$~fm and three different
temporal lattice extensions $N_t=4, 6, 8$. The physical temperature of the
system is set by its temporal extension as
\begin{equation}
  T = \frac{1}{N_t a}.
\end{equation}
In this way the lattice parameters we used correspond to the physical
temperature range of $1.7 T_c < T < 5 T_c$.

To understand the nature of the transition it is crucial to consider the
spatial volume dependence of spectral and wave function statistics. For this
reason we also repeated some simulations on different spatial volumes. The
spatial linear size of the boxes we used were all in the range $2 \,
\mathrm{fm}~\leq~L~\leq~6 \, \mathrm{fm}$.  The details of the parameters of
our ensembles are summarized in Table \ref{tab:sim_parameters}.

\begin{table}
 \begin{ruledtabular}
 \begin{tabular}{lcccccrr}
     &  $T$(MeV) & $a$(fm) & $N_s$  & $N_t$ & Nconf & Nevs \\ \hline
 A1 &  263       & 0.125   & 24    & 6    & 430   & 512 \\
 A2 &            &         & 36    &      & 420   & 256 \\ 
 B  &  300       & 0.082   & 32    & 8    & 434   & 256 \\
 C1 &  394       & 0.125   & 16    & 4    &1622   & 512 \\
 C2 &            &         & 24    &      &1600   & 512 \\
 C3 &            &         & 32    &      & 900   & 512 \\
 C4 &            &         & 48    &      & 604   & 128 \\
 D1 &  401       & 0.082   & 24    & 6    & 440   & 512 \\
 D2 &            &         & 36    &      & 440   & 256 \\
 E  &  397       & 0.062   & 32    & 8    & 593   & 256 \\ 
 F  &  530       &         &       & 6    & 420   & 512 \\ 
 G  &  601       & 0.082   & 24    & 4    & 396   & 512 \\  
 H  &  794       & 0.062   & 32    & 4    & 417   & 512 \\
 \end{tabular}
 \end{ruledtabular}
 \caption{\label{tab:sim_parameters} The parameters of the simulations; the
   temperature, the lattice spacing, the spatial and temporal box size, the
   inverse gauge coupling, the number of configurations and the number of
   Dirac eigenvalues computed on each configuration.}
\end{table}

Since low Dirac modes can potentially be slow modes of the system we also
checked for autocorrelations. In a long run performed on the finest lattice we
found that the autocorrelation time for the smallest Dirac modes is definitely
smaller than ten trajectories. To be on the safe side on the finest lattice
the configurations that we used for spectrum calculations were always
separated by thirty trajectories. Even on the coarsest lattice configurations
were separated by ten trajectories.

\section{Anderson transition in the Dirac spectrum}
     \label{se:AtitDs}


Previously it was seen that in the high temperature deconfining phase of
$SU(2)$ Yang-Mills theory there is a transition in the staggered Dirac
spectrum from localized low modes to delocalized modes higher up in the
spectrum \cite{Kovacs:2010wx}. The question we ask here is whether such a
transition also occurs in real QCD with light dynamical quarks. This is a
non-trivial question since light quarks suppress low Dirac modes through the
fermion determinant in the functional integral and in the quenched case it is
exactly the lowest Dirac modes that are localized.

The hallmark of a transition from localized to delocalized modes in terms of
spectral statistics is a change from Poisson statistics to random matrix
statistics in the spectrum. Intuitively speaking, localized modes are such
because they cannot mix with other modes; typical gauge field fluctuations do
not mix them. Localized modes close in the spectrum are peaked at spatially
distant locations and they are sensitive only to gauge field fluctuations
there. As a result the corresponding eigenvalues are statistically independent
and the level spacings obey Poisson statistics. Delocalized modes, in
contrast, are mixed by typical gauge field fluctuations. Gauge field
fluctuations change several delocalized modes together which introduces
delicate correlations in the spectrum and as a result the eigenvalue
statistics is described by Random Matrix Theory (RMT).

The simplest way to detect a transition in the spectrum from Poisson to RMT
statistics is to consider the so called unfolded level spacing
distribution. Unfolding is essentially a local rescaling of the eigenvalues to
have unit spectral density throughout the spectrum. We did the unfolding by
ordering all eigenvalues in the given ensemble and replacing them with their
rank order normalized by the total number of configurations. On a few
ensembles we also checked unfolding by using local spline approximations to
the spectral density but there was no discrepancy between the two methods of
unfolding.

Since the unfolded level spacing distribution (ULSD) is known analytically for
both the Poisson and the RMT statistics it can be easily used to distinguish
between the two cases.  For Poisson statistics the ULSD is a simple
exponential,
$$P\msub{Poisson}(s)=\exp(-s).$$
In the RMT case the unfolded level spacing distribution depends on the
universality class which in the case of staggered fermions in the fundamental
representation of the $SU(3)$ gauge group is the Chiral Unitary Ensemble
(ChUE) \cite{Verbaarschot:2000dy}. The corresponding ULSD is very precisely
approximated by the chiral unitary  Wigner surmise,
\begin{equation}
P\msub{ChUE}(s) = \dfrac{32}{\pi^{2}}s^{2}\cdot
                       \exp\left(-\dfrac{4}{\pi}s^{2}\right).
  \label{eq:wigner_chu}
\end{equation}
 
\begin{figure}
\begin{center}
\begin{tabular}{cc}
\includegraphics[width=0.50\columnwidth,keepaspectratio]{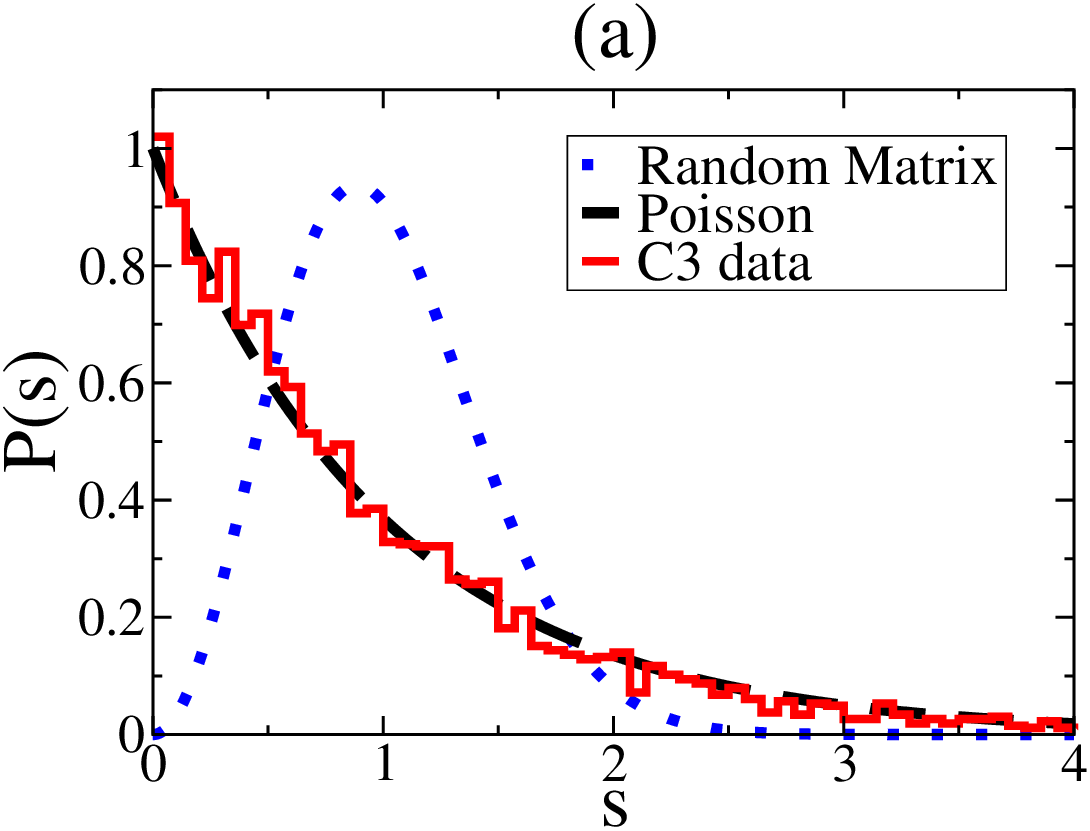} & 
\includegraphics[width=0.50\columnwidth,keepaspectratio]{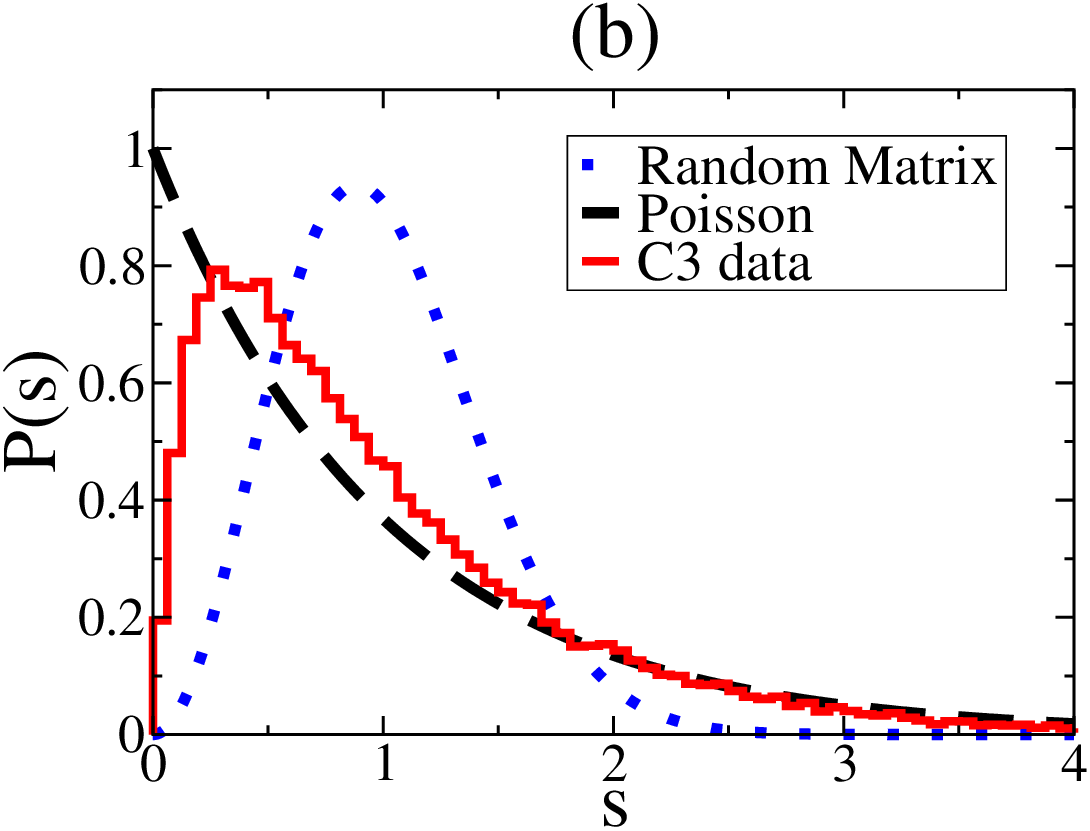} \\
\includegraphics[width=0.50\columnwidth,keepaspectratio]{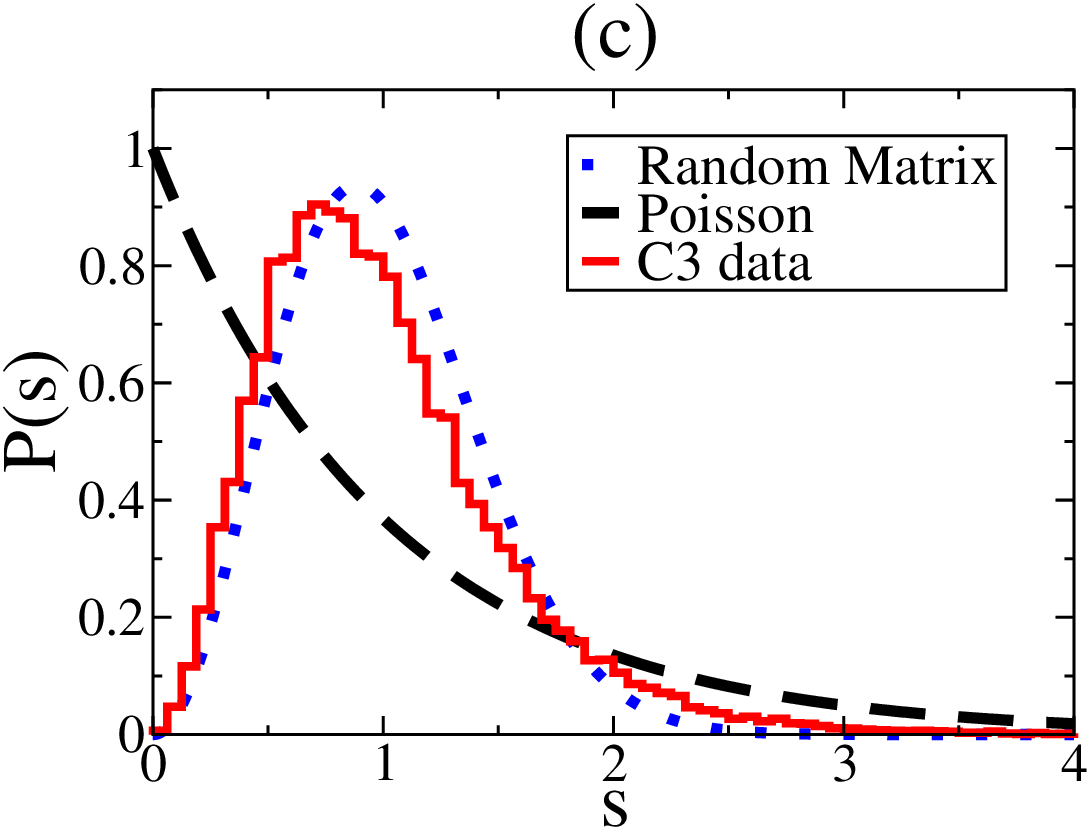} &
\includegraphics[width=0.50\columnwidth,keepaspectratio]{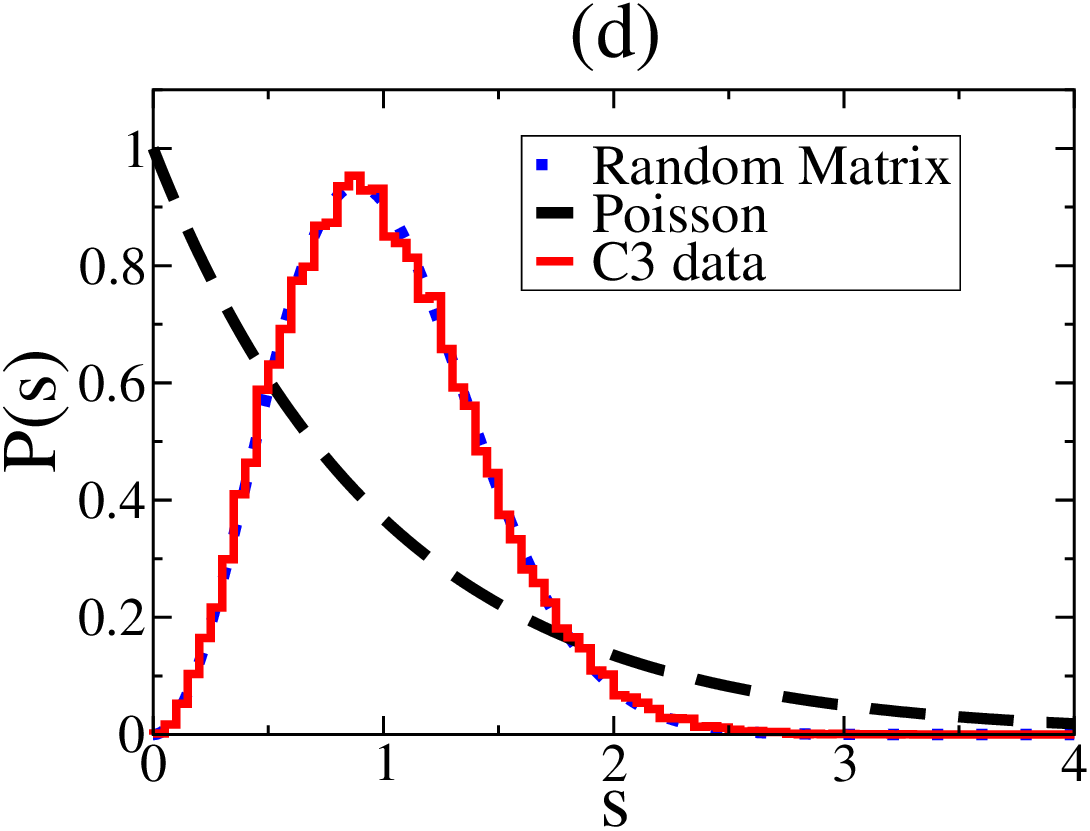} \\
\end{tabular}
\end{center}
\caption{\label{fig:ulsd}The unfolded level spacing distribution in
  different regions of the spectrum of ensemble C3. The figures correspond to
  the spectral regions $0.15 \leq \lambda a \leq 0.19$ (a), $0.29 \leq \lambda
  a \leq 0.32$ (b), $0.34 \leq \lambda a \leq 0.35$ (c) and $0.375 \leq
  \lambda a \leq 0.385$ (d). The dashed line indicates the exponential
  distribution corresponding to the localized (Poisson) case and the dotted
  line indicates the chiral unitary Wigner surmise expected in the delocalized
  (RMT) case.}
\end{figure}

To demonstrate the transition in the spectrum in Fig.~\ref{fig:ulsd} we
plot the ULSD in different regions of the spectrum of ensemble C3; $0.15 \leq
\lambda a \leq 0.19$ (a), $0.29 \leq \lambda a \leq 0.32$ (b), $0.34 \leq
\lambda a \leq 0.35$ (c) and $0.375 \leq \lambda a \leq 0.385$ (d). We also
indicate in the same Figure the distributions corresponding to the Poisson
(localized) and the RMT (delocalized) case. The transition from localized
modes at the edge of the spectrum to delocalized modes in the bulk can be
clearly seen. This shows that light dynamical fermions do not change the
picture observed in the quenched case earlier and the transition also occurs in
QCD with quarks of physical masses.

\subsection{Analogy with the Anderson transition}

The transition in the spectrum from localized to delocalized modes is
reminiscent of the Anderson metal-insulator transition occurring in conducting
crystalline solids when impurities are introduced \cite{Anderson:1958vr}. In
that case in the presence of impurities single electron Bloch states along the
band edge turn into localized states. In three dimensions if the impurity
concentration is not too high states at the band center can still remain
delocalized. The boundary between localized and delocalized states is known as
the mobility edge. Increasing the density of impurities pushes the mobility
edge further towards the center of the band until all the states in the band
become localized. When the mobility edge passes the Fermi energy and the Fermi
energy gets into the delocalized part of the spectrum the system has a
vanishing zero-temperature conductivity. The states that can be excited are
all non-conducting localized states.

It has been conjectured that the finite temperature QCD transition might be
similar to the Anderson transition \cite{GarciaGarcia:2004hi}. Further
indications to support this picture were obtained from instanton liquid
\cite{GarciaGarcia:2005vj} and lattice QCD simulations
\cite{GarciaGarcia:2006gr}. 

We now sketch the analogy between the spectrum of the one electron Hamiltonian
in disordered media and that of the QCD Dirac operator. Due to the symmetries
of the QCD Dirac operator its spectrum is symmetric with respect to the real
axis. In the continuum and also in the case of the staggered lattice Dirac
operator, the one we use here, the spectrum is purely imaginary. That is if
the quark mass is zero, otherwise the quark mass provides a trivial real part
to all the eigenvalues. In the chiral limit (zero quark mass) the spectral
density at zero is proportional to the chiral condensate, the order parameter
of spontaneous chiral symmetry breaking \cite{Banks:1979yr}. At high enough
temperature chiral symmetry is restored and the spectral density at zero
vanishes. In that case the spectrum has a so called ``soft edge'' and there
might even be a gap around zero in the spectrum
\cite{Farchioni:1999ws,Damgaard:2000cx}. This ``edge'' of the spectrum at the
low end is analogous to the band edge in condensed matter systems.

In the condensed matter literature numerical studies of the Anderson
transition usually concentrate on the band center and locate the critical
disorder when states at the center become localized. As we will see, in the
case of QCD the location of the mobility edge is controlled by the temperature
and there is no analog of the band center since the spectral density continues
to be non-zero all the way up to the cut-off scale. Therefore, unlike in most
of the condensed matter literature, here we do not attempt to determine the
critical disorder where all delocalized states disappear but rather study how
the mobility edge changes with the temperature. This approach is not
completely unknown in the condensed matter literature either
\cite{Siringo:1998}.

\subsection{Second moment of the unfolded level spacing distribution}
   \label{ssec:smulsd}

As can be seen in Fig.~\ref{fig:ulsd} the unfolded level spacing distribution
changes in the spectrum from the exponential to the Wigner surmise in a
continuous fashion.  There does not appear to be a sharp mobility edge,
$\lambda_c$, separating localized and delocalized states. Even in the case of
Anderson transitions, however, a sharp transition is expected only in the
thermodynamic limit when the spatial size of the system, $N_s$, goes to
infinity.  This is completely analogous to second order phase transitions
where a truly divergent correlation length and a sharp phase transition can
only be observed in infinite systems. In principle any physical quantity that
changes in a well defined way from the localized to the delocalized regime can
be used to define a transition point in the spectrum. If there is a sharp
transition in the thermodynamic limit then non-analytic behavior should appear
in all these quantities at a given point, $\lambda_c$, in the spectrum. In
what follows we will look for a quantity that can be used to define the
transition point.

\begin{figure}
\begin{center}
\includegraphics[width=\columnwidth,keepaspectratio]{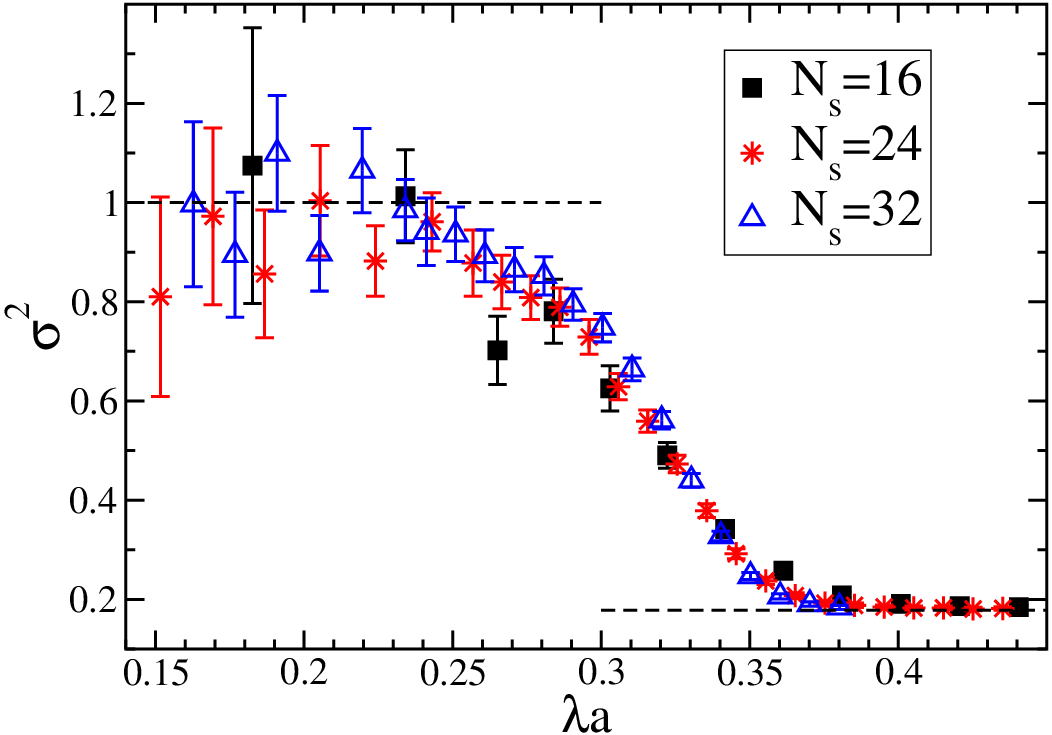}
\caption{\label{fig:ulsdvar} Variance of the unfolded level spacing
  distribution across the transition in the spectrum for ensembles C1, C2 and
  C3 that differ only in their spatial volume. The dashed horizontal lines at
  1 and 0.178 indicate the expected limits in the localized (Poisson) and
  delocalized (RMT) regime.}
\end{center}
\end{figure}

\begin{figure}
\begin{center}
\includegraphics[width=\columnwidth,keepaspectratio]{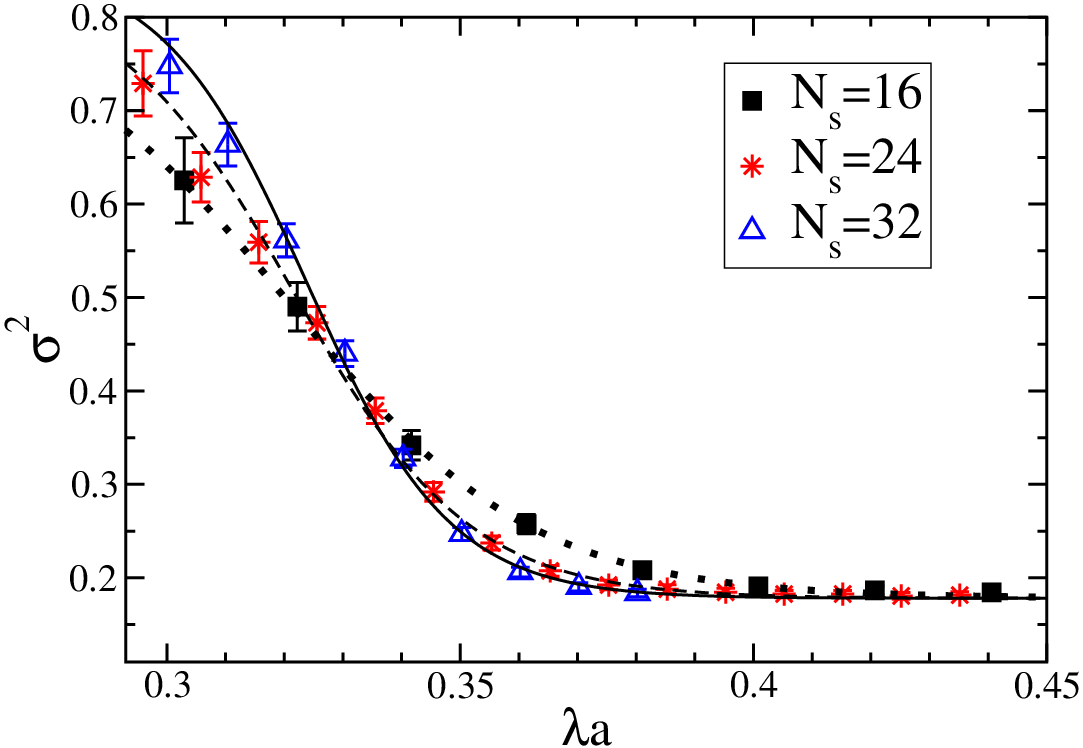}
\caption{\label{fig:ulsdvar_zoom} The same as Fig.~\ref{fig:ulsd} but zooming
  in on the transitions region. The dotted, dashed and the continuous curve
  indicate the fits of the form Eq.\ (\ref{eq:tanh}) to the $N_s=16, 24$ and
  $32$ data respectively.}
\end{center}
\end{figure}

Since the first moment of the unfolded level spacing distribution is unity by
construction, the simplest quantity to consider is its second moment or
variance.  In the localized case the level spacings are exponentially
distributed and the variance is 
\begin{equation}
 \sigma\msub{s}^2 = \langle s^2 \rangle - \langle s \rangle^2=1,
\end{equation}
while in the delocalized regime the second moment of the distribution of
Eq.\ (\ref{eq:wigner_chu}) can be analytically determined to be
$\sigma\msub{s}^2=\frac{3\pi}{8}-1$. In Fig.~\ref{fig:ulsdvar} we plot how the
variance changes in the spectrum throughout the transition region. The three
data sets correspond to ensembles C1-C3 with three different spatial volumes
but otherwise identical parameters. It is apparent that with increasing volume
the transition becomes sharper. To define a finite volume pseudocritical point
in the spectrum we locate the inflection point of the curve
$\sigma^2(\lambda)$. To this end we use the three-parameter fitting Ansatz
\begin{equation}
  \sigma^2(\lambda) = A\left\{ 1 - \tanh(B(\lambda-C)) \right\} + 
          \frac{3\pi}{8}-1.
    \label{eq:tanh}
\end{equation}
This form ensures the correct limit for large $\lambda$ and yields good fits
starting already from $\lambda a = 0.2, 0.25$ i.e. already from below the
transition point. Using this Ansatz the inflection point can be easily seen to
be at $\lambda\msub{c} a=C$ and the slope of the curve there is $AB$. These
two parameters are largely independent of the starting point of the fit as
long as it starts at smaller values of $\lambda$ than where the inflection
point occurs. The location of the inflection point turns out to be also
independent of the volume yielding pseudocritical points $C=\lambda_ca =
0.321(4), 0.322(1), 0.324(2)$ for the spatial sizes $N_s=16, 24, 32$. The
slope at the inflection point is $AB=7.6(5), 10.96(36), 13.8(5)$ for the three
different spatial sizes and it scales roughly proportionally to the linear
spatial size of the box. This indicates that there might be a genuine sharp
transition in the thermodynamic limit. The sharpening of the transition with
the volume can be better seen in Fig.~\ref{fig:ulsdvar_zoom} where we plot
again the variance of the unfolded level spacing distribution but zoom in on
the transition region and also show the fitted curves with the above described
parameters.

\subsection{Eigenvector statistics}

Besides the spectrum the spatial profile of the corresponding eigenvectors
also contains important information concerning their localization
properties. A quantity that is widely used in this context is the inverse
participation ratio (IPR) defined as \cite{wegner}
\begin{equation}
 P_\psi = \sum_x \vert \psi(x) \vert^{4},
   \label{eq:IPR}
\end{equation}
where $\psi$ is an eigenvector normalized as
\begin{equation}
  \sum_x \vert \psi(x) \vert^2 = 1.
\end{equation}
The qualitative physical meaning of the IPR can be easily seen by noting that
an eigenmode that spreads uniformly in a four-volume $v$ and is zero everywhere
else has $\mbox{IPR}=1/v$. The IPR$^{-1}$ thus measures the volume
occupied by an eigenmode. Alternatively one can use the participation ratio,
$\mbox{PR}=\mbox{IPR}^{-1} \cdot V$, where $V$ is the total volume of the
system. This measures the fraction of the total volume occupied by the
eigenmode. 

The behavior of the IPR and the PR in the thermodynamic limit can be used to
distinguish between localized and delocalized modes. By definition a part of
the spectrum consists of localized modes if their average IPR remains
finite as the volume goes to infinity. This also implies that their average PR
vanishes in the thermodynamic limit. In contrast delocalized modes have
vanishing IPR which usually implies non-vanishing PR. 

\begin{figure}
\begin{center}
\includegraphics[width=\columnwidth,keepaspectratio]{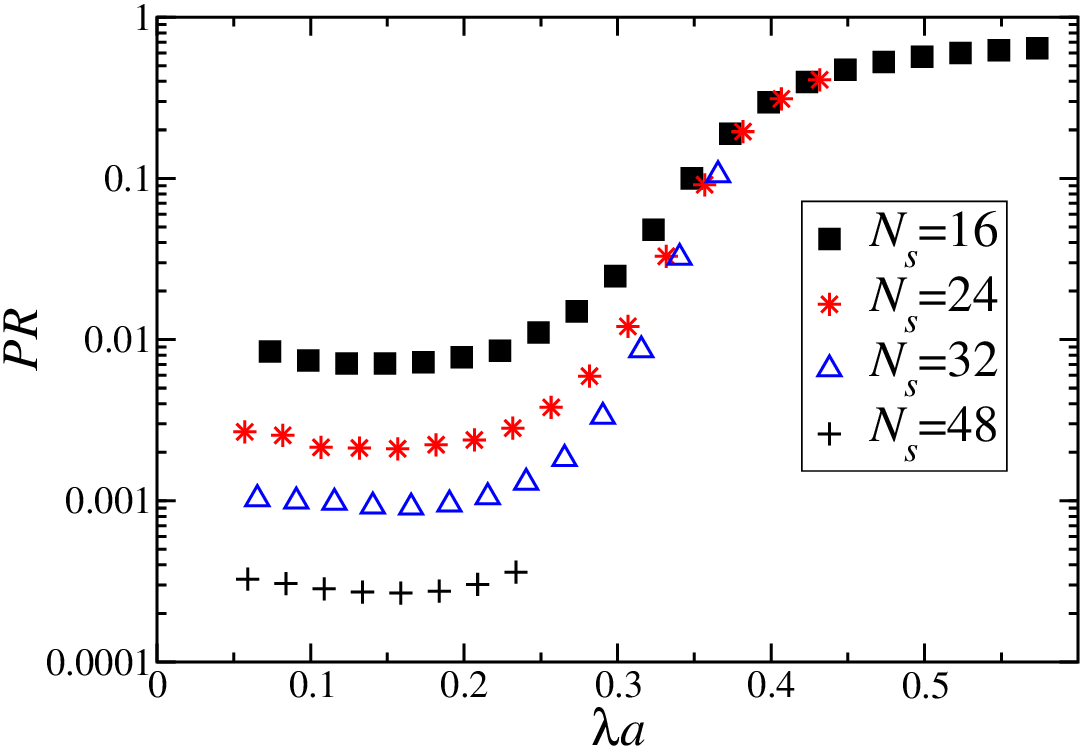}
\caption{\label{fig:particip} Average $PR$ of the eigenvectors for lattices
  with the same temperature and lattice spacing but with four different aspect
  ratios.$N_{t}=4$}
\end{center}
\end{figure}

In Fig.~\ref{fig:particip} we plot how the average participation ratio of
eigenmodes changes throughout the spectrum on ensembles C1-C4. The four data
sets correspond to different spatial sizes but otherwise identical
parameters. It is clearly seen that the average PR for the low eigenmodes
decreases as the volume is increased; the average PR here tends to zero in the
thermodynamic limit. This suggests that these eigenmodes are localized. They
fill a vanishing fraction of the total box volume in the thermodynamic
limit. Higher up in the spectrum the average PR becomes a volume independent
constant of order unity, which means that these eigenmodes are
delocalized. They fill a non-zero fraction of the total box volume in the
thermodynamic limit.

\section{Continuum limit and renormalization}
    \label{se:Clar}

We saw that on all the ensembles that we considered there is a critical point
(``mobility edge''), $\lambda_c$, in the Dirac spectrum that separates
localized and delocalized eigenmodes. If the lowest part of the spectrum
becomes localized that can have a dramatic effect on long distance correlators
of quark operators and the masses associated with them. This happens because
in the spectral decomposition of the quark propagator, $(D+m)^{-1}$, each
eigenmode is weighted by the inverse of the corresponding
eigenvalue. Therefore the lowest eigenmodes receive the largest weight. On the
other hand, eigenmodes localized to a distance scale $l$ have negligible
contribution to correlators at distance scales much larger than $l$.

To assess the physical implications of localization in QCD we are thus lead to
study two questions;
\begin{enumerate}
 \item What is the distance scale $l$ on which the lowest eigenmodes are
   localized?
 \item How far up in the spectrum $(\lambda\msub{c})$ are the modes localized?
\end{enumerate}
Both of these questions have to be considered in the continuum limit as the
lattice spacing $a\rightarrow 0$.

\subsection{Localization length}

\begin{figure}
\begin{center}
\includegraphics[width=\columnwidth,keepaspectratio]{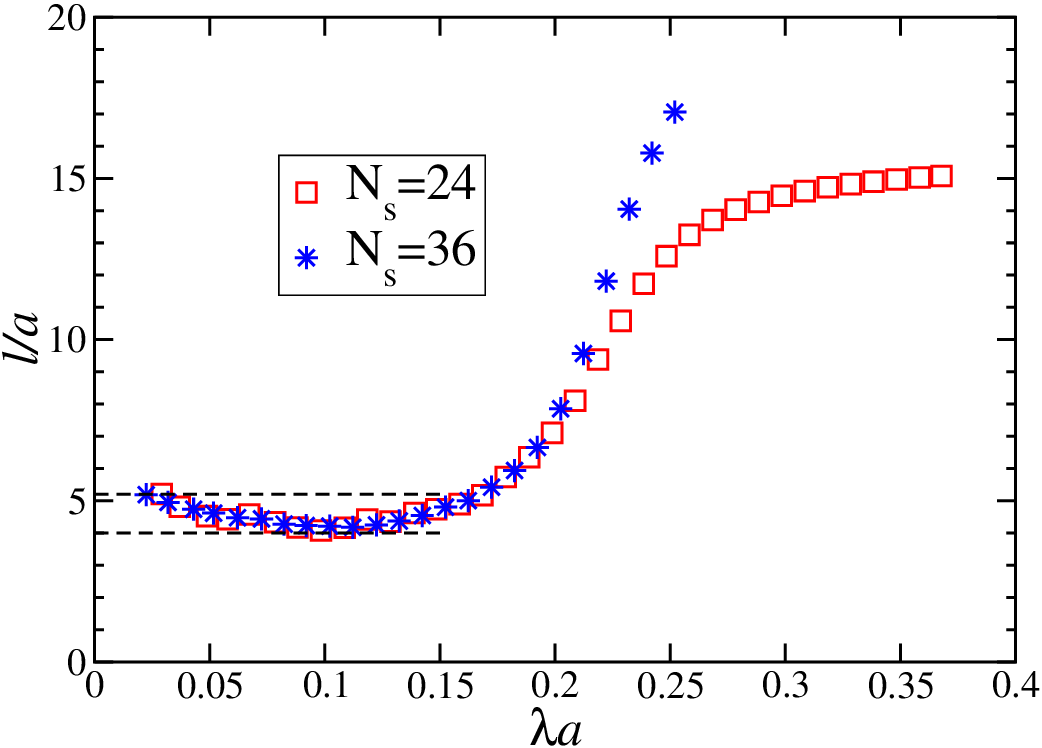}
\caption{\label{fig:lvslambda} The localization length (Eq.\ \ref{eq:l}) of
  eigenmodes along the spectrum for ensembles D1 and D2 differing only in
  their spatial volumes. Both $l$ and the location in the spectrum, $\lambda$
  are given in lattice units. Statistical errors are not shown as they are
  smaller than the size of the symbols.}
\end{center}
\end{figure}

To get a rough estimate of the localization scale we can consider the inverse
participation ratio (IPR) defined by Eq.\ \ref{eq:IPR}. Since the IPR scales
like the inverse four-volume of the region where the given mode is spread out a
good measure of the localization scale is provided by the quantity
\begin{equation}
 l = a \cdot \langle \mbox{IPR}^{-\frac{1}{4}} \rangle,
   \label{eq:l}
\end{equation}
where the average is understood over all the eigenmodes in a given region of
the spectrum. In principle $l$ varies through the spectrum but it does not
change too much within the region of localized modes. To illustrate that, in
Fig.~\ref{fig:lvslambda} we show in two representative cases how $l$
changes through the spectrum. As can be seen in the Figure for the lowest part
of the spectrum the localization length is almost constant and is independent
of the spatial volume. In this region there might be a small dip in $l$. The
dip is generally more pronounced at lower temperatures and/or on finer
lattices but on our ensembles the total variation of $l$ in the localized
regime never exceeds 20\%. At some point in the spectrum $l$ starts to
increase sharply and becomes strongly volume dependent. This is the beginning
of the transition to delocalized states.

In what follows we define the localization length of localized modes with the
following simple procedure. The localized eigenmodes all have $l$'s between
the bottom of the dip and the $l$ of the very lowest eigenmodes. This interval
for ensembles D1 and D2 is indicated by the two dashed horizontal lines in
Fig.~\ref{fig:lvslambda}. The central value we quote for $l$ is always the
center of this band and the uncertainty is half the width of the band.
Compared to that, statistical errors are always negligible.

\begin{figure}
\begin{center}
\includegraphics[width=\columnwidth,keepaspectratio]{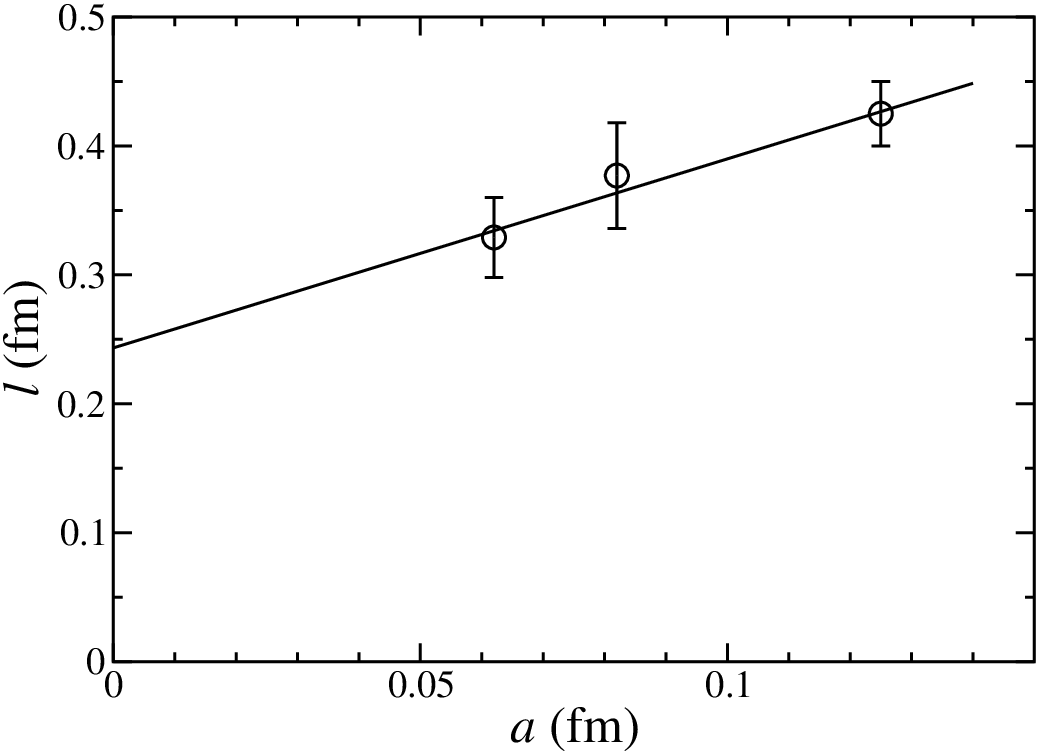}
\caption{\label{fig:lvsa} The localization length of the localized modes as a
  function of the lattice spacing. The three points correspond to the same
  physical temperature of about 400MeV (ensembles C, D and E).}
\end{center}
\end{figure}

Having a well defined measure of the localization length for localized modes
we can now look at how it depends on the lattice spacing. Since, as we will
see, $l$ also depends on the temperature, we choose to compare ensembles C, D
and E which are almost exactly at the same physical temperature of about
400~MeV. Since $l$ does not depend on the spatial volume we omit the numbers
from the ensemble labels here.  In Fig.~\ref{fig:lvsa} we plot the
localization length as a function of the lattice spacing for these three
ensembles.

To guide the eye we also included a linear fit to the data. Even if the
quality of our data does not allow a proper continuum extrapolation it can be
safely concluded that the localization length measured in physical units does
not increase in the continuum limit and it is not larger than a few tenth of a
Fermi at this temperature.

\begin{figure}
\begin{center}
\includegraphics[width=\columnwidth,keepaspectratio]{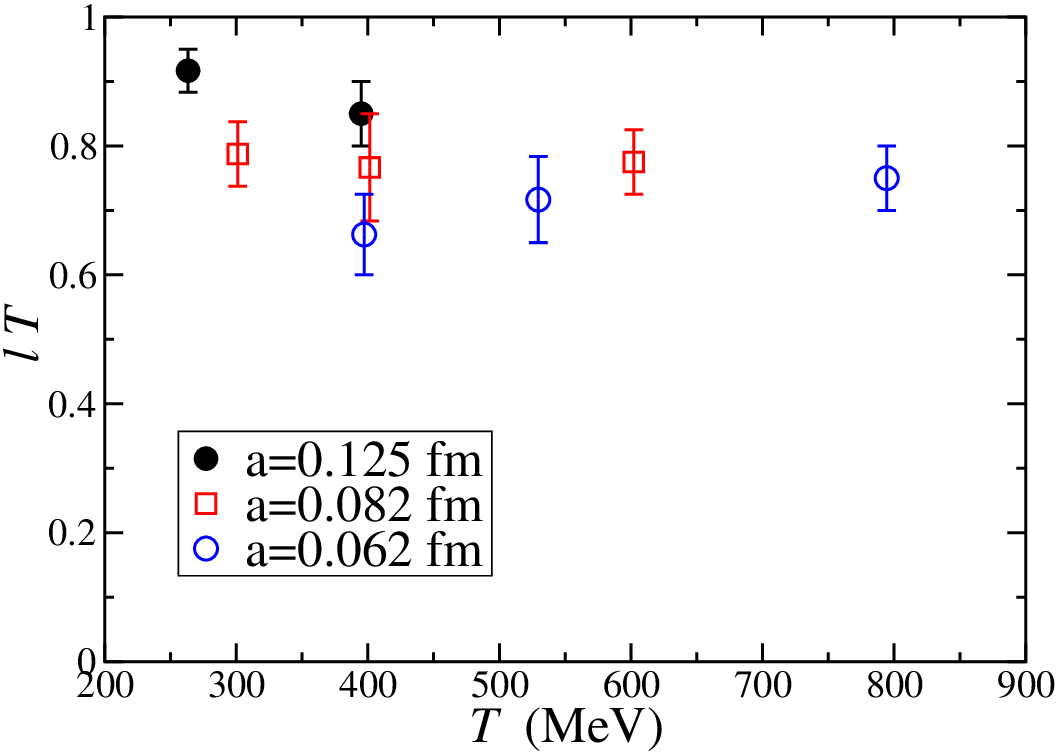}
\caption{\label{fig:lvsT} The localization length in units of the inverse
  temperature as a function of the physical temperature. The different symbols
represent data obtained at different values of the lattice spacing.}
\end{center}
\end{figure}

It is also instructive to see how the localization length compares to the most
important length scale in the problem, the inverse temperature or in other
words the temporal size of the box. In Fig.~\ref{fig:lvsT} we plot the
localization length in units of the inverse temperature for all the
ensembles. Different symbols represent data sets corresponding to different
values of the lattice spacing. There is a slight trend of the finer lattices
producing more localized low modes as can be seen also in
Fig.~\ref{fig:lvsa}. The important point is that the lowest modes always
appear to be localized on or below the scale of the inverse temperature in the
whole range of temperatures studied here.

\subsection{Renormalization and temperature dependence of $\lambda_c$}

We have seen that the lowest eigenmodes spread out only to a distance scale
below the inverse temperature. Therefore these modes cannot contribute to
quark propagation on length scales larger than that. As far as hadronic
correlators are considered above that distance scale the system behaves as if
it had a gap of order $\lambda_c$. The influence of this effective gap on the
physics depends on how $\lambda_c$, the critical point in the spectrum, scales
in the continuum limit.

In Subsection \ref{ssec:smulsd} we showed how to determine the critical point
using the second moment of the unfolded level spacing distribution. This
procedure yields the critical point $\lambda_c a$ in dimensionless lattice
units for each temperature and lattice spacing. In what follows we give a
proposal for defining the continuum limit of this quantity. Since the critical
point is effectively a gap for quark modes capable of propagating to long
distances it plays a role similar to the quark mass that also introduces a
gap. For this reason we expect $\lambda_c$ to be renormalized in exactly the
same way as the quark mass and the ratio of $\lambda_c a$ to the bare light
quark mass $m\msub{ud} a$ should have a proper continuum limit. Moreover this
quantity measures the relative size of the effective gap for delocalized modes
and the gap for all modes provided by the quark mass.

Another, perhaps more formal, argument showing that $\lambda_c/m\msub{ud}$ has
a well-defined continuum limit is as follows. On the one hand, the
pseudoscalar meson correlator $\langle P(x) P(0) \rangle$ is proportional to
the matrix elements of the square of the Dirac propagator,
$[(D^\dagger+m)(D+m)]^{-1}$. On the other hand, for asymptotically large
temporal separations $t$
\begin{equation}
    \frac{1}{V\msub{s}} \sum_{\vec{x}} \langle P(t,\vec{x}) P(0) \rangle =
    C\msub{PP} \mbox{e}^{m_\pi t},
\end{equation} 
where $V\msub{s}$ is the spatial volume and $C\msub{PP}$ is related to the
pion decay constant, $f_\pi$, as \cite{Aubin:2004fs}
\begin{equation}
   C\msub{PP} m\msub{ud}^2 = \frac{f_\pi^2}{V\msub{s}} m_\pi^3.
\end{equation}
The right-hand side of this equation is a well-defined physical quantity and
thus so is the left-hand side. Since $C\msub{PP}$ is proportional to the
inverse quark propagator squared, it is also proportional to the inverse of
the Dirac eigenvalues squared. Thus
\begin{equation}
 m\msub{ud}^2 \lambda\msub{c}^{-2} \propto C\msub{PP} m\msub{ud}^2
\end{equation}
also has a well defined continuum limit. For a more general discussion of
similar issues with Wilson fermions see also \cite{DelDebbio:2005qa} and
\cite{Giusti:2008vb}.

\begin{figure}
\begin{center}
\includegraphics[width=\columnwidth,keepaspectratio]{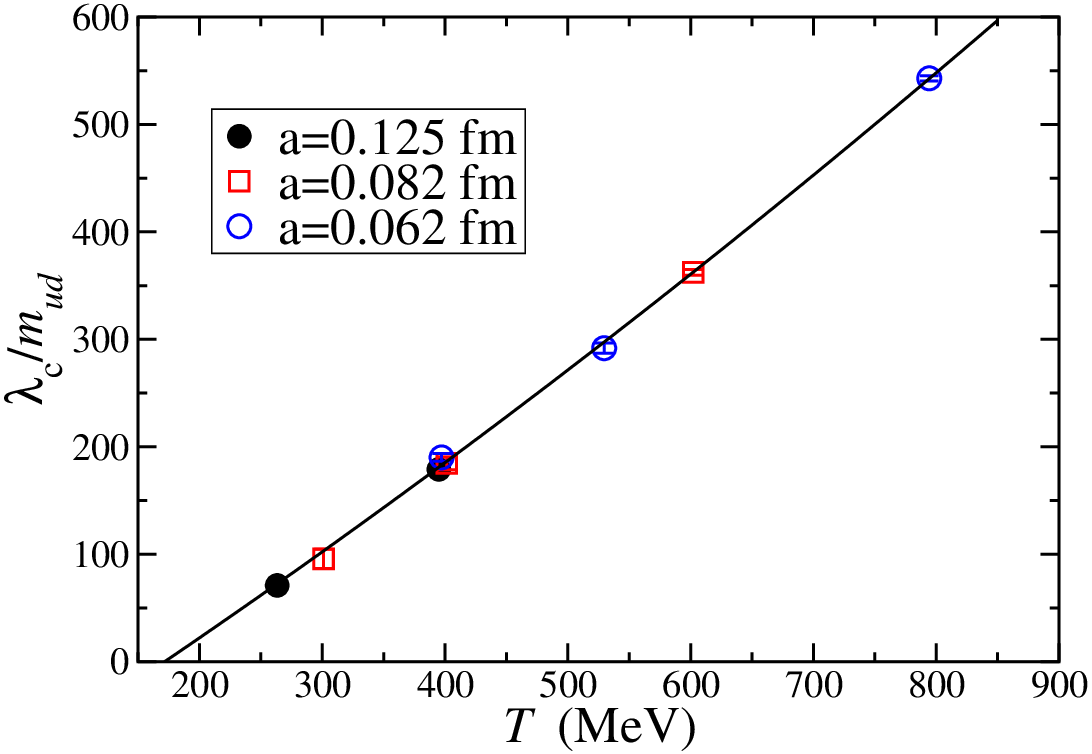}
\caption{\label{fig:lcpmvsT} The temperature dependence of the renormalized
  critical point $\frac{\lambda\msub{c}}{m\msub{ud}}$. The continuous line is
  a quadratic polynomial fit to all the data. The different symbols correspond
to data obtained using lattices of different coarseness.}
\end{center}
\end{figure}

Data sets C, D and E where we have data at the same temperature for all three
lattice spacings indicate that the lattice spacing dependence in the quantity
$\frac{\lambda\msub{c}}{m\msub{ud}}$ is comparable to its uncertainty at a
given lattice spacing. Therefore in Fig.~\ref{fig:lcpmvsT} we plot the
temperature dependence of this quantity for all lattice spacings in the same
figure. The different types of symbols, corresponding to data obtained from
lattices of different coarseness, all lie on a smooth curve confirming that
scaling violations in this quantity are small. The lowest order polynomial fit
to all the data that yields an acceptable chi-squared ($\chi^2=1.3$) is second
order and extrapolates to $T=$170MeV at
$\frac{\lambda\msub{c}}{m\msub{ud}}=0$. This value is consistent with the
known location of the finite temperature cross-over in QCD
\cite{Borsanyi:2010bp, Bazavov:2011nk} below which chiral symmetry is broken
and no localized modes are expected to be present. This provides a further
consistency check of our results. 

\begin{figure}
\begin{center}
\includegraphics[width=\columnwidth,keepaspectratio]{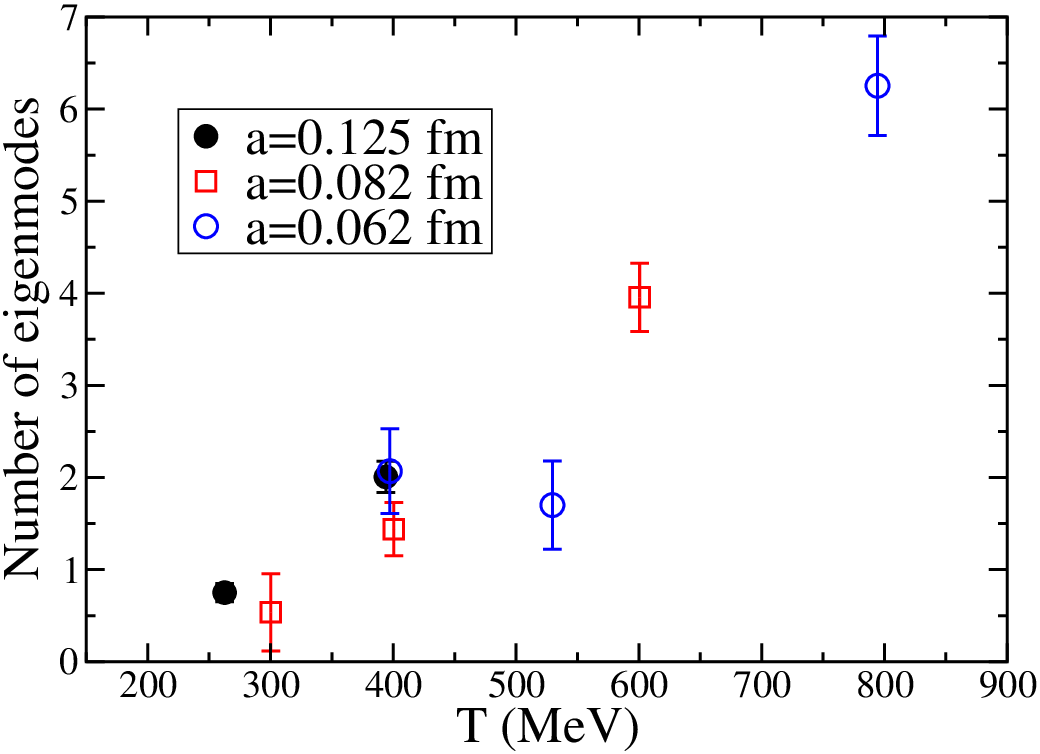}
\caption{\label{fig:locnum} The number of localized (below $\lambda\msub{c}$)
  eigenmodes per cubic Fermi as a function of the temperature. Different
  symbols refer to simulations at different values of the lattice spacing.}
\end{center}
\end{figure}

Another physically interesting quantity is the number density of localized
modes. This quantity is also practically important if one aims at detecting
localization from spectral statistics. In order to be able to observe Poisson
statistics in the level spacing distribution the volume of the system has to
be large enough to accommodate several localized eigenmodes per
configuration. To see what that means practically in Fig.~\ref{fig:locnum} we
plotted the average density of eigenmodes below $\lambda\msub{c}$ as a
function of the temperature. Again we collect data obtained at different values
of the lattice spacing in the same Figure. Apparently the number density of
localized modes decreases sharply as the transition temperature is approached
from above. In practical terms that means that in order to have for instance
about twenty localized modes per configuration the aspect ratio of the boxes
has to be kept between 4-6 in the temperature range considered here.

Finally we would like to point out a potential technical difficulty in
studying the level spacing distribution. It is caused by the pairing of
staggered Dirac eigenvalues affecting the lowest part of the spectrum when the
spatial volume is small. It comes about because in the continuum limit
staggered fermions describe four fermion flavors and the spectrum is expected
to become fourfold degenerate. Therefore approaching the continuum limit this
degeneracy starts to be formed and peculiar correlations will appear among
members of the would-be continuum quartets.

It turns out that already at the lattice spacings we used here the stout
smearing of gauge fields coupled to the fermions bring the spectrum close
enough to the continuum behavior that such extra correlations for the lowest
eigenvalues can be detected. This is first manifested in pairwise attraction
between consecutive eigenvalues which distorts the Poisson statistics
\cite{Kovacs:2011km}.  The splitting between members of the doublets, however,
does not depend appreciably on the volume while the separation of the doublets
(eventually quartets) should be inversely proportional to the spectral density
and thus to the volume. Therefore this distortion of the level spacing
statistics is a finite volume effect (c.f.\ also
\cite{Halasz:1996sb}). Moreover it only affects the lowest part of the
spectrum where the spectral density is very small. Since the spatial volumes
we used are large enough the results we presented here are not affected by the
staggered degeneracy. On the other hand, a more systematic study of the level
spacing statistics of the very lowest staggered Dirac modes might reveal
interesting facts about how the expected degeneracy is formed in the continuum
limit.

\section{Conclusions}
   \label{se:C}

In the present paper we argued that in QCD above the finite temperature
cross-over the lowest eigenmodes of the quark Dirac operator are
localized. The spatial localization length is set by the inverse temperature
(see Fig.~\ref{fig:lvsT}) with eigenmodes becoming more ``squeezed'' at
higher temperature. At the same time when the temperature increases the
mobility edge, separating localized and delocalized modes, is also pushed
higher up in the spectrum.

This high temperature behavior of the low Dirac modes has to be contrasted
with the situation at low temperature below the cross-over. In that case
chiral symmetry is spontaneously broken and a finite density of eigenmodes
extends down all the way to zero. As a result, the statistics of the lowest
Dirac modes is described by random matrix theory. QCD is thus a remarkably
rich theory. The low-end of the Dirac spectrum can exhibit both possible
extremes of spectral statistics; maximally mixed modes with RMT statistics
below the transition and completely independent eigenmodes with Poisson
spectral statistics at high temperature.

The localization of the lowest Dirac modes can dramatically suppress hadronic
correlators at high temperatures. In the eigenmode expansion of the quark
propagator the lowest part of the spectrum receives the largest relative
weight. If, as we saw, these modes are localized they cannot propagate quarks
to long distances resulting in a suppression of long-distance correlators. As
can be seen in Fig.~\ref{fig:lcpmvsT} the mobility edge below which all states
are localized moves steeply up with increasing temperature. Already at $2T_c$
it is two orders of magnitude larger than the bare light quark mass. This
mechanism might help to explain the steep rise of screening masses above $T_c$
seen in lattice simulations \cite{Cheng:2010fe}.

An interesting question is how exactly the spectral and wave function
statistics change through the mobility edge and whether there is any
universality in how these quantities interpolate between the localized and
delocalized regime. In particular we would like to check whether the scale
invariance of the inverse participation ratio distribution observed in
Anderson transitions \cite{logP2} also occurs in the QCD transition. We hope
to return to this question in a future publication. As far as the Poisson to
RMT transition in the spectral statistics is concerned there does not seem to
be a universal understanding of how it happens in general but there are
several proposals that might provide further insight
\cite{Schierenberg:2012ut}-\cite{NishigakiPRE}.

It would also be interesting to know what physical mechanism drives the
transition. Is it possible to identify some physical objects in the gauge
field background that are responsible for the appearance of localized
modes? In a previous paper some evidence was found that there is a correlation
between localized modes and local fluctuations of the Polyakov loop
\cite{Bruckmann:2011cc}. On the other hand it was also argued there that
uncorrelated instantons cannot play a significant role in this mechanism as
their density is too low for that. It is, however, still possible that
instanton molecules or ``bions'' have to do with localization
\cite{Anber:2011gn}. A better understanding of the physical mechanism behind
localization in QCD could possibly shed some more light on the finite
temperature chiral and deconfining transition.

\section*{Acknowledgments}

TGK is supported by the Hungarian Academy of Sciences under ``Lend\"ulet''
grant No.\ LP2011-011. Both authors acknowledge partial support by the EU
Grant (FP7/2007-2013)/ERC No. 208740. We thank the Budapest-Wuppertal group
for allowing us to use their computer code for generating the lattice
configurations. Finally we also thank S. D.\ Katz and D.\ N\'ogr\'adi for
discussions.


\begin{thebibliography}{99}

\bibitem{Banks:1979yr}
  T.~Banks and A.~Casher,
  Nucl.\ Phys.\  B {\bf 169}, 103 (1980).

\bibitem{Verbaarschot:2000dy}
  J.~J.~M.~Verbaarschot and T.~Wettig,
  Ann.\ Rev.\ Nucl.\ Part.\ Sci.\  {\bf 50}, 343 (2000)
  [arXiv:hep-ph/0003017].

\bibitem{DeGrand:2009et} 
  T.~DeGrand,
  arXiv:0906.4543 [hep-lat];
   %
  T.~DeGrand,
  Phys.\ Rev.\ D {\bf 80}, 114507 (2009)
  [arXiv:0910.3072 [hep-lat]].

\bibitem{DelDebbio:2010ze} 
  L.~Del Debbio and R.~Zwicky,
  Phys.\ Rev.\ D {\bf 82}, 014502 (2010)
  [arXiv:1005.2371 [hep-ph]].

\bibitem{Patella:2012da} 
  A.~Patella,
  arXiv:1204.4432 [hep-lat].

\bibitem{Cheng:2011ic} 
  A.~Cheng, A.~Hasenfratz and D.~Schaich,
  Phys.\ Rev.\ D {\bf 85}, 094509 (2012)
  [arXiv:1111.2317 [hep-lat]];
  A.~Hasenfratz, A.~Cheng, G.~Petropoulos and D.~Schaich,
  arXiv:1207.7162 [hep-lat].

\bibitem{Farchioni:1999ws} 
  F.~Farchioni, P.~de Forcrand, I.~Hip, C.~B.~Lang and K.~Splittorff,
  Phys.\ Rev.\ D {\bf 62}, 014503 (2000)
  [hep-lat/9912004].

\bibitem{Damgaard:2000cx} 
  P.~H.~Damgaard, U.~M.~Heller, R.~Niclasen and K.~Rummukainen,
  Nucl.\ Phys.\ B {\bf 583}, 347 (2000)
  [hep-lat/0003021].

\bibitem{Bohigas:1983er} 
  O.~Bohigas, M.~J.~Giannoni and C.~Schmit,
  Phys.\ Rev.\ Lett.\  {\bf 52}, 1 (1984).

\bibitem{Anderson:1958vr} 
  P.~W.~Anderson,
  Phys.\ Rev.\  {\bf 109}, 1492 (1958).

\bibitem{Halasz:1995vd} 
  A.~M.~Halasz and J.~J.~M.~Verbaarschot,
  Phys.\ Rev.\ Lett.\  {\bf 74}, 3920 (1995)
  [hep-lat/9501025].

\bibitem{Pullirsch:1998ke} 
  R.~Pullirsch, K.~Rabitsch, T.~Wettig and H.~Markum,
  Phys.\ Lett.\ B {\bf 427}, 119 (1998)
  [hep-ph/9803285].

\bibitem{GarciaGarcia:2004hi} 
  A.~M.~Garcia-Garcia and K.~Takahashi,
  Nucl.\ Phys.\ B {\bf 700}, 361 (2004)
  [cond-mat/0403557].

\bibitem{GarciaGarcia:2005vj} 
  A.~M.~Garcia-Garcia and J.~C.~Osborn,
  Nucl.\ Phys.\ A {\bf 770}, 141 (2006)
  [hep-lat/0512025].

\bibitem{GarciaGarcia:2006gr} 
  A.~M.~Garcia-Garcia and J.~C.~Osborn,
  Phys.\ Rev.\ D {\bf 75}, 034503 (2007)
  [hep-lat/0611019].

\bibitem{Kovacs:2009zj} 
  T.~G.~Kovacs,
  Phys.\ Rev.\ Lett.\  {\bf 104}, 031601 (2010)
  [arXiv:0906.5373 [hep-lat]].

\bibitem{Kovacs:2010wx}
  T.~G.~Kovacs and F.~Pittler,
  Phys.\ Rev.\ Lett.\  {\bf 105}, 192001 (2010)
  [arXiv:1006.1205 [hep-lat]].


\bibitem{Gavai:2008xe} 
  R.~V.~Gavai, S.~Gupta and R.~Lacaze,
  Phys.\ Rev.\ D {\bf 77}, 114506 (2008)
  [arXiv:0803.0182 [hep-lat]];
     %
  V.~G.~Bornyakov, E.~V.~Luschevskaya, S.~M.~Morozov, M.~I.~Polikarpov,
  E.~-M.~Ilgenfritz and M.~Muller-Preussker,
  Phys.\ Rev.\ D {\bf 79}, 054505 (2009)
  [arXiv:0807.1980 [hep-lat]].

\bibitem{deForcrand:2006my} 
  P.~de Forcrand,
  AIP Conf.\ Proc.\  {\bf 892}, 29 (2007)
  [hep-lat/0611034].

\bibitem{Aoki:2005vt} 
  Y.~Aoki, Z.~Fodor, S.~D.~Katz and K.~K.~Szabo,
  JHEP {\bf 0601}, 089 (2006)
  [hep-lat/0510084].
 
\bibitem{Borsanyi:2010cj} 
  S.~Borsanyi, G.~Endrodi, Z.~Fodor, A.~Jakovac, S.~D.~Katz, S.~Krieg,
  C.~Ratti and K.~K.~Szabo, 
  JHEP {\bf 1011}, 077 (2010)
  [arXiv:1007.2580 [hep-lat]].

\bibitem{Siringo:1998} 
  F.~Siringo and G.~Piccitto,
  J.\ Phys.\ A A {\bf 31}, 5981 (1998).

\bibitem{wegner} 
  F.~Wegner,
  Z.\ Phys.\ {\bf B36} (1980) 209.


\bibitem{Aubin:2004fs} 
  C.~Aubin {\it et al.}  [MILC Collaboration],
  Phys.\ Rev.\ D {\bf 70}, 114501 (2004)
  [hep-lat/0407028].

\bibitem{DelDebbio:2005qa} 
  L.~Del Debbio, L.~Giusti, M.~Luscher, R.~Petronzio and N.~Tantalo,
  JHEP {\bf 0602}, 011 (2006)
  [hep-lat/0512021].

\bibitem{Giusti:2008vb} 
  L.~Giusti and M.~Luscher,
  JHEP {\bf 0903}, 013 (2009)
  [arXiv:0812.3638 [hep-lat]].

\bibitem{Borsanyi:2010bp} 
  S.~Borsanyi {\it et al.}  [Wuppertal-Budapest Collaboration],
  JHEP {\bf 1009}, 073 (2010)
  [arXiv:1005.3508 [hep-lat]].

\bibitem{Bazavov:2011nk} 
  A.~Bazavov, T.~Bhattacharya, M.~Cheng, C.~DeTar, H.~T.~Ding, S.~Gottlieb,
  R.~Gupta and P.~Hegde {\it et al.}, 
  Phys.\ Rev.\ D {\bf 85}, 054503 (2012)
  [arXiv:1111.1710 [hep-lat]].

\bibitem{Kovacs:2011km} 
  T.~G.~Kovacs and F.~Pittler,
  PoS LATTICE {\bf 2011}, 213 (2011)
  [arXiv:1111.3524 [hep-lat]].

\bibitem{Halasz:1996sb} 
  A.~M.~Halasz, T.~Kalkreuter and J.~J.~M.~Verbaarschot,
  Nucl.\ Phys.\ Proc.\ Suppl.\  {\bf 53}, 266 (1997)
  [hep-lat/9607042].

\bibitem{Cheng:2010fe} 
M.~Cheng, S.~Datta, A.~Francis, J.~van der Heide,
  C.~Jung, O.~Kaczmarek, F.~Karsch and E.~Laermann {\it et al.},
  Eur.\ Phys.\ J.\ C {\bf 71}, 1564 (2011)
  [arXiv:1010.1216 [hep-lat]].

\bibitem{logP2}
  Y.V.~Fyodorov and A.D.~Mirlin, 
  Phys.\ Rev.\ B {\bf 51}, 13403 (1995);
    %
 F.~Evers and A.D.Mirlin,
 Phys.\ Rev.\ Lett.\ {\bf 84}, 3690 (2000);
     %
 A.~Mildenberger, F.~Evers and A.D.~Mirlin,
 Phys.\ Rev.\ B {\bf 66}, 033109 (2002);
    %
A.D.~Mirlin and F.~Evers,
Phys.\ Rev.\ B {\bf 62}, 7920 (2000).

\bibitem{Schierenberg:2012ut} 
  S.~Schierenberg, F.~Bruckmann and T.~Wettig,
  Phys.\ Rev.\ E {\bf 85}, 061130 (2012)
  [arXiv:1202.3925 [math-ph]].


\bibitem{Bruckmann:2011cc} 
  F.~Bruckmann, T.~G.~Kovacs and S.~Schierenberg,
  Phys.\ Rev.\ D {\bf 84}, 034505 (2011)
  [arXiv:1105.5336 [hep-lat]].

\bibitem{Nishigaki:2012jw} 
  S.~M.~Nishigaki,
  arXiv:1208.3878 [hep-lat];
  %
  S.~M.~Nishigaki,
  arXiv:1208.3452 [hep-lat].

\bibitem{NishigakiPRE}
  S.~M.~Nishigaki,
  Phys.\ Rev.\  E {\bf 58}, R6915 (1998); 
     %
  S.~M.~Nishigaki, 
  Phys.\ Rev.\  E {\bf 59}, 2853 (1999). 

\bibitem{Anber:2011gn} 
  M.~M.~Anber, E.~Poppitz and M.~Unsal,
  JHEP {\bf 1204}, 040 (2012)
  [arXiv:1112.6389 [hep-th]];
    %
  M.~Unsal,
  arXiv:1201.6426 [hep-th];
     %
  E.~Poppitz, T.~Schaefer and M.~Unsal,
  arXiv:1205.0290 [hep-th].





\end{thebibliography}
\end{document}